\documentclass[]{raa} 
\usepackage{graphicx,times}
\usepackage{natbib}
\newcommand{\ra}[1]{\renewcommand{\arraystretch}{#1}}

\providecommand{\bm}[1]{\mbox{\boldmath$#1$\unboldmath}}
\def\etal{et~al}
\def\aec{atmospheric extinction coefficients}
\def\ae{atmospheric extinction}
\def\nsb{night sky brightness}

\begin{document}

\title{
Atmospheric Extinction Coefficients and 
Night Sky Brightness At the Xuyi Observational Station 
}

\volnopage{ {\bf ????} Vol.\ {\bf ?} No. {\bf XX}, 000--000}

\setcounter{page}{1}
   \author{ H.-H Zhang \inst{1} 
       \and X.-W. Liu \inst{1,2} 
       \and H.-B. Yuan \inst{2,3}
       \and H.-B. Zhao  \inst{4}
       \and J.-S. Yao   \inst{4}
       \and H.-W. Zhang \inst{1}
       \and M.-S. Xiang \inst{1}}

\institute{ Department of Astronomy, Peking 
            University, Beijing 100871, P.~R.~China; \\
          { \it ~~~~x.liu@pku.edu.cn} \\
      \and  Kavli Institute for Astronomy and Astrophysics,  
	    Peking University, Beijing 100871, P.~R.~China; \\
      \and  LAMOST Fellow; \\
      \and  Purple Mountain Observatory, CAS, Nanjing 210008, P.~R.~China. \\}

\date{Received [201?] [month] [day]; accepted [201?] [month] [day] }

\abstract{ 
We present measurements of the optical broadband \aec~and the \nsb~at the Xuyi
Observational Station of Purple Mountain Observatory (PMO). The measurements
are based on CCD imaging data taken in the Sloan Digital Sky Survey $g$, $r$
and $i$ bands with the Xuyi 1.04/1.20\,m Schmidt Telescope for the Xuyi Schmidt
Telescope Photometric Survey of the Galactic Anti-center (XSTPS-GAC), the
photometric part of the Digital Sky Survey of the Galactic Anti-center
(DSS-GAC). The data were collected in more than 130 winter nights from 2009 to
2011.  We find that the atmospheric extinction coefficients for the $g$, $r$
and $i$ bands are 0.70, 0.55 and 0.38 mag/airmass, respectively, based on
observations taken in several photometric nights.  The \nsb~determined from
images of good quality has median values of 21.7, 20.8 and 20.0
mag/arcsec$^{\rm 2}$ and reaches 22.1, 21.2 and 20.4
mag/arcsec$^{\rm 2}$ under the best observing conditions for the $g$, $r$ and
$i$ bands, respectively.  The relatively large extinction coefficients compared
with other good astronomical observing sites are mainly due to the relatively
low elevation (i.e. 180\,m) and high humidity of the Station. 
}

   \authorrunning{H.-H. Zhang \etal. }    
   \titlerunning{Atmospheric Extinction Coefficients and Night Sky Brightness at the Xuyi Station}  
   \maketitle

\section{Introduction}    
\label{sect:intro}

The \aec~and \nsb~are key parameters characterizing the quality of an
astronomical observational site, in addition to the seeing and the number of
clear night. The \ae~is the brightness reduction of celestial objects as their
light passes through the earth atmosphere. 
\cite{Parrao2003} points out
that precise atmospheric extinction determinations are needed not only for
stellar photometry but also for any sort of photometry, spectroscopy,
spectrophotometry and imaging whenever accurate, absolute and well-calibrated
photometric measurements are required for the derivation of physical parameters
in the studies of galaxies, nebulae, planets, and so forth. Precise
determinations of the \ae~ultimately determine the scientific value of the
telescope data. The \nsb~is caused by the scattered starlight, the airglow,
zodiacal light and the artificial light pollution from the nearby cities. It
is the main noise sources of ground-based astronomical observations. The
\nsb~limits the detection depth of a telescope -- the lower the \nsb, the
fainter the stars that could be detected and the more astronomical information
that could be collected. 

All good astronomical sites around the world have been subject to comprehensive
studies of their atmospheric extinction properties and \nsb~measurements. For
example, 
\cite{Kris1987} 
and 
\cite{Kris1990} 
study the
Mauna-Kea site, the host of the CFHT, the Keck 1 \& 2, the Subaru telescopes as
well as the future Thirty Meter Telescope (TMT) under development. 
\cite{Burki1995} 
and 
\cite{Mattila1996} 
study the La Silla site, and their results are
used to refer the characteristics of the Las Campanas site that hosts the Giant
Magellan Telescope (GMT). 
\cite{Tokov2006} 
models the optical
atmospheric turbulence for the Cerro Pachon site where has been selected for
the Large Synoptic Survey Telescope (LSST). The \ae~and \nsb~of the La Palma,
site of the Gran Telescopio Canarias (GTC) and of a number of smaller
telescopes have been studied by 
\cite{Garcia2010} 
and 
\cite{Benn1998}, respectively. 
\cite{Pat2011} measures the \ae~of the Cerro Paranal
site of the Very Large Telescopes (VLT). 
Parrao \etal. (2003) studies the
atmospheric extinction of the San Pedro M\'{a}rtir (SPM) site. 

In those studies above different methods of measuring the \ae~and \nsb~have
been proposed and the results are discussed in term of affecting factors such
as the aerosols, the water vapor content and the airglow, zodiacal light and
light pollution. 
\cite{Kris1990} and \cite{Kris1997} also study and confirm
the effects of the solar activities on the night sky measurements. 
\cite{Kris1991} constructs a model for the brightness of moonlight and 
\cite{Garstang1989} investigates the sky brightness caused by the night glows. 
\cite{Hogg2001}  
studies the atmospheric extinction of the Apache Point Observatory (APO)
and describe a near real-time extinction monitor instruments that has
significantly improved the photometric calibration of the Sloan Digital Sky
Survey (SDSS). 
\cite{Burke2010}
proposes a method for precise determinations
of the atmospheric extinction by studying the absorption signatures of
different atmospheric constituents, which might be applied for future
ground-based surveys such as the LSST. Real-time and accurate measurements of
the night sky atmospheric properties will be of utter importance for future
ground-based surveys. 

The Digital Sky Survey of the Galactic Anti-center 
(DSS-GAC; Liu \etal. 2012, in preparation), is a spectroscopic and photometric survey targeting millions
of stars distributed in a contiguous sky area of about 3,500 sq.deg.~centered
on the Galactic Anti-center. The spectroscopic component of the DSS-GAC, the
Guo Shou Jing Telescope (LAMOST) Spectroscopic Survey of the Galactic
Anti-center (GSJTSS-GAC) will secure optical spectra for a statistically
complete sample over three million stars of all colors and spectral types,
whereas its photometric component, the Xuyi Schmidt Telescope Photometric
Survey of the Galactic Anti-center (XSTPS-GAC) surveys the area in the SDSS
$g$, $r$ and $i$ bands with wide field CCD using the Xuyi 1.04/1.20\,m Schmidt
Telescope in order to provide the spectroscopic target input catalogs for the
GSJTSS-GAC. XSTPS-GAC was initiated in 2009 and completed in March 2011. In
total, over 20,000 images have been collected over more than 140 nights. The
survey reaches a limiting magnitude of 19 (10$\sigma$) in all three bands.
Approximately 100 millions stars have been detected and cataloged in $i$~band,
and approximately half the number in $g$~band.

The Xuyi Schmidt Telescope is located on a small hill about 35\,km from the
Xuyi country in middle-east of China with an elevation of about 180\,m above
the sea level, similar to the elevation of Fowler's Gap in Australia, a
potential site for a Cherenkov telescope 
\citep{Hampf2011}. The atmospheric
extinction and \nsb~of the Xinglong station where the LAMOST is located have
been studied by 
\cite{YanHJ2000} and 
\cite{Liu2003}, respectively. In this
work, we present measurements of the \aec~and \nsb~of the Xuyi Station using
the data collected for the XSTPS-GAC. The data have been analyzed using the
traditional methods.  The telescope characteristics and the basic data
reduction steps are briefly outlined in Section 2. Section 3 presents
measurements of the \aec. In Section 4, we describe statistical measurements of
the \nsb~and discuss potential effects of the moon phase/brightness and the
contribution of the Galactic disk on the night sky brightness measurements.
Finally, the conclusions follow in Section 5.  

\section{Observations and Data Reduction}
\label{sect:Obs}

The Xuyi Schmidt Telescope is a traditional ground-based refraction-reflective
telescope with a diameter of 1.04/1.20\,m. It was equipped with a thinned
4096$\times$4096 CCD camera, yielding a 1.94$^{\rm o} \times $ 1.94$^{\rm o}$
effective field-of-view (FoV) at a sampling of 1.705\,arcsec per pixel
projected on the sky. The CCD quantum efficiency, at the cooled working
temperature of $-$103.45$^{\rm o}$C, has a peak value of 90 percent in the blue
and remains above 70 percent even to wavelengths as long as 8,000\,{\AA}. The
XSTPS-GAC was carried with the SDSS $g$, $r$ and $i$ filters. The current work
presents measurements of the Xuyi \aec~and the \nsb~in the three SDSS filters
based on the images collected for the XSTPS-GAC. 

The XSTPS-GAC images in total a sky area of 6,000 deg$^2$ centered on the
Galactic Anti-center, from 3$h$ to 9$h$ in right ascension (RA) and from
$-$10$^{\rm o}$ to $+$60$^{\rm o}$ in declination (DEC), plus an extension
about 900 deg$^2$ area toward the M31/M33 area. The sky coverage is shown in
Fig.\,2. Most observations were carried out in dark or grey nights of good
photometric quality. A small fraction of the observations were taken under
bright lunar conditions and in those cases, the angular distances between the
field centers and the Moon were kept at greater than 60\,deg.  An integration
time of 90\,sec was used for all exposure, with a readout time of approximately
43\,sec in dual-channel slow readout mode. The field center stepped in RA by
half the Fov (i.e. 0.97\,deg.), leading to two exposures for a given point of
the sky. Two stripes of scan of adjacent declinations overlapped by
approximately 0.04\,deg. Normally, two stripes of fields of adjacent
declinations were scanned in a given night. The stripes cross the Galactic disk
from the south to the north, a fact that we have found useful to quantify the
effects of the bright Galactic disk on the measurement of night sky brightness.
With this specific combination of integration time and scanning strategy, the
movement of the telescope pointing was minimized in a given night, ensuring a
maximum uniformity of the survey. To facility the global flux calibration, a
few "Z" stripes of fields that straddled between the fields of "normal" stripes
were also observed. Finally, in the course of the survey, in order to measure
the \aec, several photometric nights were picked out and used to carry out
repeated exposures of a few pre-selected fixed fields at different zenith
distances, with a time interval of half an hour. 

\begin{table}[t]\bc
\begin{minipage}[]{120mm}
\caption{Fields used for extinction coefficient measurements and the results}
\end{minipage}
\small
\ra{1.3}
\begin{tabular}{c c c c c c}
\hline\hline\noalign{\smallskip}
  date 	& field$^{a}$ & filter & $k_1$ & $k_2$ & $k_c$ \\ \hline
2010/01/06 & 073824+2134 & $g$~ & 0.70717 $\pm$ 0.01077 & -0.02571 $\pm$ 0.00106 &  0.02824 $\pm$  0.01320 \\
2010/01/14 & 073824+2134 & $g$~ & 0.73861 $\pm$ 0.01271 & -0.02760 $\pm$ 0.00145 &  0.03436 $\pm$  0.00746 \\
     & 073824+2134$^{b}$ & $g$~ & 0.64923 $\pm$ 0.06397 & -0.10747 $\pm$ 0.00373 &  0.00774 $\pm$  0.00000 \\
2010/10/04 & 011800+3000 & $g$~ & 0.51588 $\pm$ 0.01714 & -0.03288 $\pm$ 0.00117 &  0.03595 $\pm$  0.01222 \\
       & 011800+3000$^b$ & $g$~ & 0.60061 $\pm$ 0.02199 & -0.01594 $\pm$ 0.00326 &  0.02165 $\pm$  0.00000 \\
           & 202600+3000 & $g$~ & 0.76053 $\pm$ 0.01180 &  0.00178 $\pm$ 0.00131 & -0.00360 $\pm$  0.00025 \\
2010/09/30 & 011800+3000 & $r$~ & 0.45208 $\pm$ 0.00773 & -0.00749 $\pm$ 0.00126 &  0.01167 $\pm$  0.00330 \\
2010/10/05 & 011800+3000 & $r$~ & 0.64922 $\pm$ 0.01528 & -0.01693 $\pm$ 0.00087 &  0.02317 $\pm$  0.00303 \\
       & 011800+3000$^b$ & $r$~ & 0.39911 $\pm$ 0.01269 &  0.00457 $\pm$ 0.00154 & -0.00870 $\pm$  0.00156 \\
           & 202600+3000 & $r$~ & 0.55799 $\pm$ 0.01338 & -0.00035 $\pm$ 0.00143 &  0.00874 $\pm$  0.00194 \\
2010/01/13 & 073824+2134 & $i$~ & 0.38335 $\pm$ 0.01579 & -0.00006 $\pm$ 0.00081 & -0.00428 $\pm$  0.00068 \\
\hline\noalign{\smallskip}
\end{tabular}
\ec
\tablenotes{a}{0.85\textwidth}{This column is a string expression of the field in a hhmmss+ddmm format.}
\tablenotes{b}{0.85\textwidth}{The second observation of the same field on descending path from the zenith. }
\end{table}

Each raw image was bias-subtracted and flat-fielded, using a super-sky-flat
(SSF) generated from all frames taken in the same filter in the same night.
There is interference fringing in the $i$ band images. However, the fringing
pattern is found to be stable for a given night, and thus can be effectively
removed by SSF fielding. Aperture and PSF photometry was then performed using a
package developed by the Beijing-Arizona- Taiwan-Connecticut (BATC) Sky Survey
\citep{Fan1996, ZhouX2001}
based on the widely used package of DAOPHOT
\citep{Stetson1987}. The astrometry was initially calibrated with the GSC2.0
\citep{Bucc2001} reference catalog, and then the plate distortion was
corrected using a 30-parameters solution based on the PPMXL 
\citep{Roser2010} reference catalogs, yielding an accuracy of about 0.1\,arcsec (Zhang
\etal., in preparation). An UBERCAl 
\citep{Padman2008} photometric calibration was achieved by
calibrating against overlapping SDSS DR8 
\citep{sdss_dr8} fields, yielding
a global photometric accuracy and homogeneity of 2-3 percent over the whole
survey area, which could be seen in  
(Yuan \etal., in preparation).

\section{Atmospheric extinction coefficients}
\label{sect:aec}

\begin{figure}{}
\centering
\includegraphics[width=140mm]{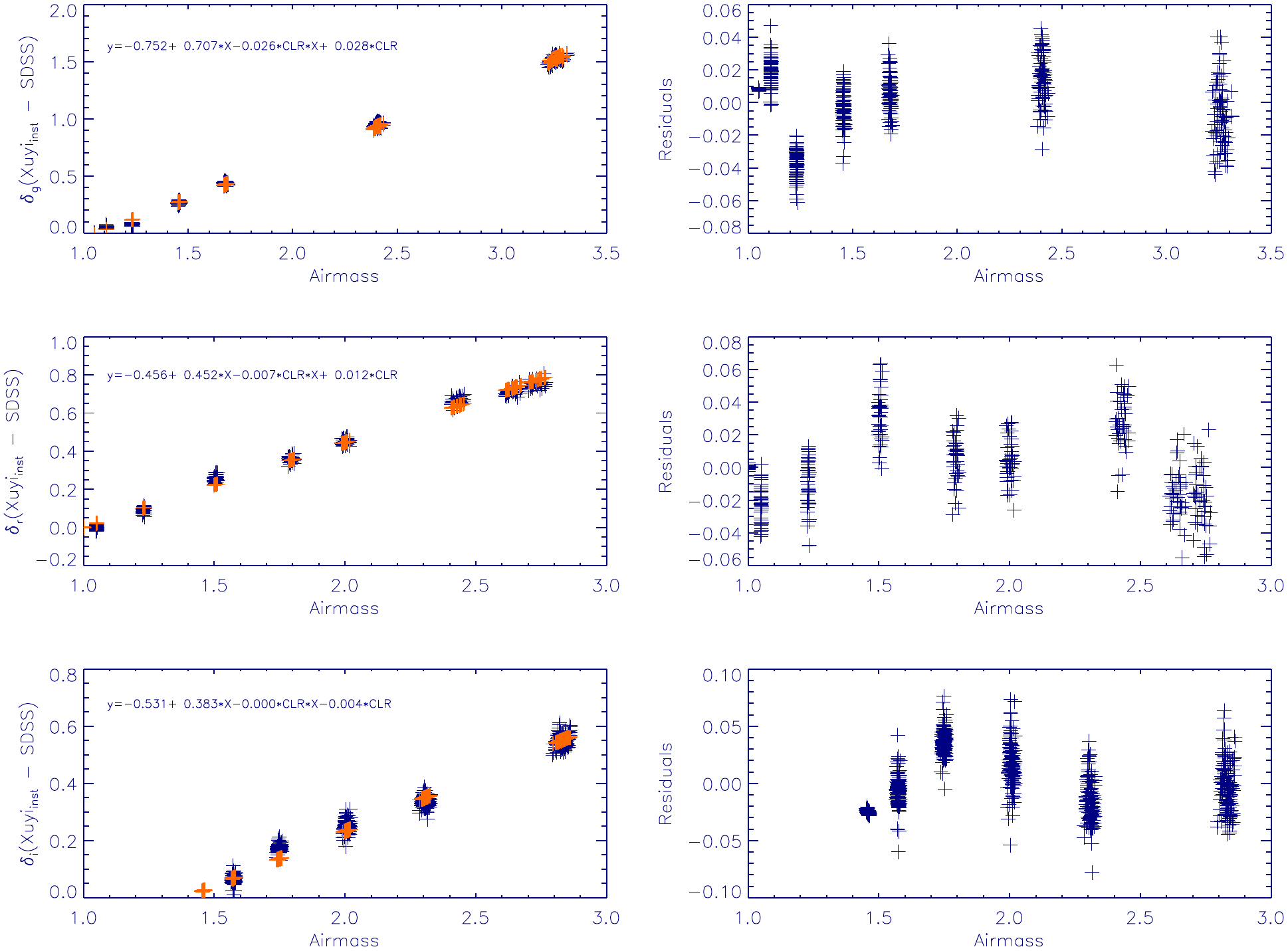}
\caption{The atmospheric extinction coefficients were estimated by fitting all
data points of all usable stars in a given field as is illustrated here for
$g$, $r$ and $i$ bands (from top to bottom), respectively. {\it Left}: The
measured instrument magnitude relative to the SDSS measurements as a function
of airmass, with observed values marked in blue pluses and fitted values in
orange diamonds. {\it Right}: The fitting residuals as a function of airmass.}
\end{figure}

In the course of the XSTPS-GAC survey, six photometric nights were picked out
in order to measure the \aec~-- three nights for the $g$ band, two nights for
the $r$ band and one night for the $i$ band. In each night, repeated
observations were carried out for one or two pre-selected fields, at different
airmass with about half hour interval between the exposures.  The observed
fields for individual nights are listed in Table 1.  In the current
observations, given the large FoV of nearly 4\,sq.deg., thousands bright,
non-saturated stars of high signal-to-noise ratios as well as good astrometric
calculation are captured in each of the individual frame of exposure, and they
can be utilized to determine the astrometric extinction coefficients. For this
purpose, we need color information of individuals stars. As such, we choose our
fields for atmospheric extinction measurements to overlap with SDSS stripes.
The SDSS source catalogs, which go deeper than ours, then provide accurate
colors for all stars utilized in the current work. Those colors are treated as
"intrinsic" outside the Earth atmosphere. However, note that there are some
slight differences between the XSTPS-GAC and SDSS filters. Amongst the series
of images taken for each of the field, measurements made on the one of the
smallest airmass were adopted as the references. 

Individual frames were scrutinized and those of poor quality were rejected.
Only isolated point sources within the central 1.0 sq.deg.~of the field and
of good signal-to-noise ratios were selected.  Specifically, we required that
the stars should have SDSS $r$ band point-spread-function (PSF) magnitudes
brighter than 17.0 and photometric errors less than 0.02. In addition, the
measurement uncertainties of the Xuyi photometry should also be less than
0.02\,mag. The stars were then examined for potential variables. The observed
magnitudes of individual stars at different values of airmass were plotted as a
function of airmass, and sources possessing 3$\sigma$ outliers to a linear
regression were rejected, as they were likely candidates of variable stars of
large variation amplitudes. 

Given the large FoV, the airmass of individual stars within the frame has to be
calculated separately. In addition, the effect of the spherical Earth needs to
be taken into account. The following formula of airmass given by 
\cite{Hardie1962} 
has been adopted:
\begin{equation}
X=\sec z -0.0018167 (\sec z -1) -0.002875(\sec z -1)^2 -0.0008083(\sec z -1)^3
\end{equation}   
where $z$ is the zenith angle in degree. The zenith angle of individual stars
can be calculated from their measured celestial coordinates, the
observational UT time and date. Then the instrument magnitude of a
star is linked to its "intrinsic" value at the smallest airmass by the
relation, 
\begin{equation}
m_{inst}=m_{0} + k_1 X + k_2 XC +k_c C +const 
\end{equation}
where $m_{inst}$ is the stellar instrument magnitude, $m_{0}$ the instrument 
magnitude of the same star at the lowest airmass, 
$k_1$ the primary extinction coefficient, $k_2$ the second-order 
extinction coefficient that relates to the color of the star, $k_c$ 
a color correction factor that accounts for the small differences between the 
SDSS and Xuyi filter systems, $C$ the intrinsic color of the star which taking to 
be the $(g-i)$ value as given by the SDSS photometry here, and finally $const$ a
constant reflecting the zero point drift of the photometric system. 

For each observed field, at least 50 stars are usable, with values of airmass
of individual exposures spanning from 1.0 to 3.0 or larger. The stars also span
a wide range of color of more than 0.5 mag. Thus more than 50 equations of the
form of Eq. (2) could be constructed. Instead of resolving them individually,
we fit all the data by a polynomial fitting in order to estimate the three
extinction coefficients, $k_1$, $k_2$ and $k_c$. The results from individual
fields are listed in Table\,1. Fig.\,1 shows three examples of the fitting
results and their residual distribution for each band. From Table\,1, we can see that, 
the second extinction coefficient $k_2$ and the color coefficient
$k_c$ shows some real variations with time and seems to be needed better constrained. 
The cause of the variations is probably related to the
changes of observing conditions, such as the humidity, winds and so on. The
primary extinction coefficients $k_1$ deduced for all three bands are large
compared to typical values of some of the best astronomical sites (Mauna Kea,
La Palama etc.), probably due to the low elevation of the Xuyi site. $k_1$ also
shows large variations from night to night. It seems that $k_1$ becomes smaller
after midnight, probably reflecting the drop of temperature and the content of
aerosol in the air. The average values of $k_1$ deduced from all observations
are 0.69, 0.50 and 0.38 for the $g$, $r$ and $i$ band, respectively. The
$i$~band coefficient  is less well determined as there was only one night
observation available. The relatively large value of the $i$~band coefficient
again possibly reflects the fact that aerosol is the dominant source of
extinction for such a low altitude site. 

There are three main sources of atmospheric extinction: the Rayleigh
scattering, ozone absorption and aerosol dust extinction. 
\cite{Bessell1990} provides empirical formula to fit the extinction of each of the three
components. The magnitudes of the Rayleigh scattering [$K_R \propto {\rm
exp}(-h_0 /8) \lambda ^{-4}$] and of the aerosol absorption [$K_A \propto {\rm
exp}(-h_0 /1.5) \lambda ^{-0.8}$] both depend on $h_0$, the altitude in km of
the site above the sea level. The low altitude of Xuyi site is clearly the main
reason for the relatively large extinction coefficients, especially that of the
$g$ band, compare to other good astronomical sites which are generally at much
higher altitudes. Given the significant variations of the extinction
coefficients from night to night, some care must be exercised if one uses
the average values presented here to reduce the whole data. 

\section{Night sky brightness}
\label{sect:nsb}

\begin{figure}{}
\centering
\includegraphics[width=49mm]{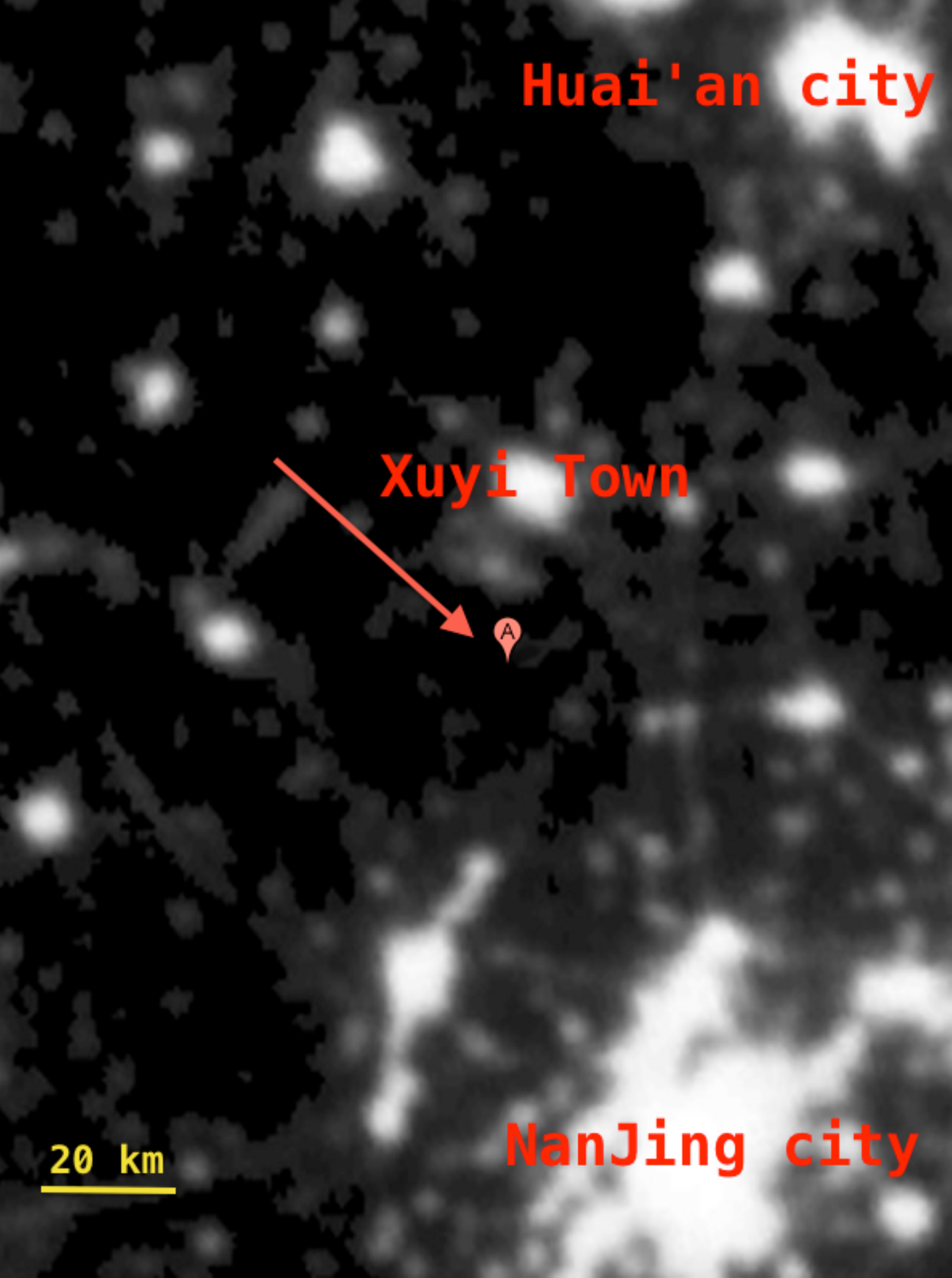}
\includegraphics[width=90mm]{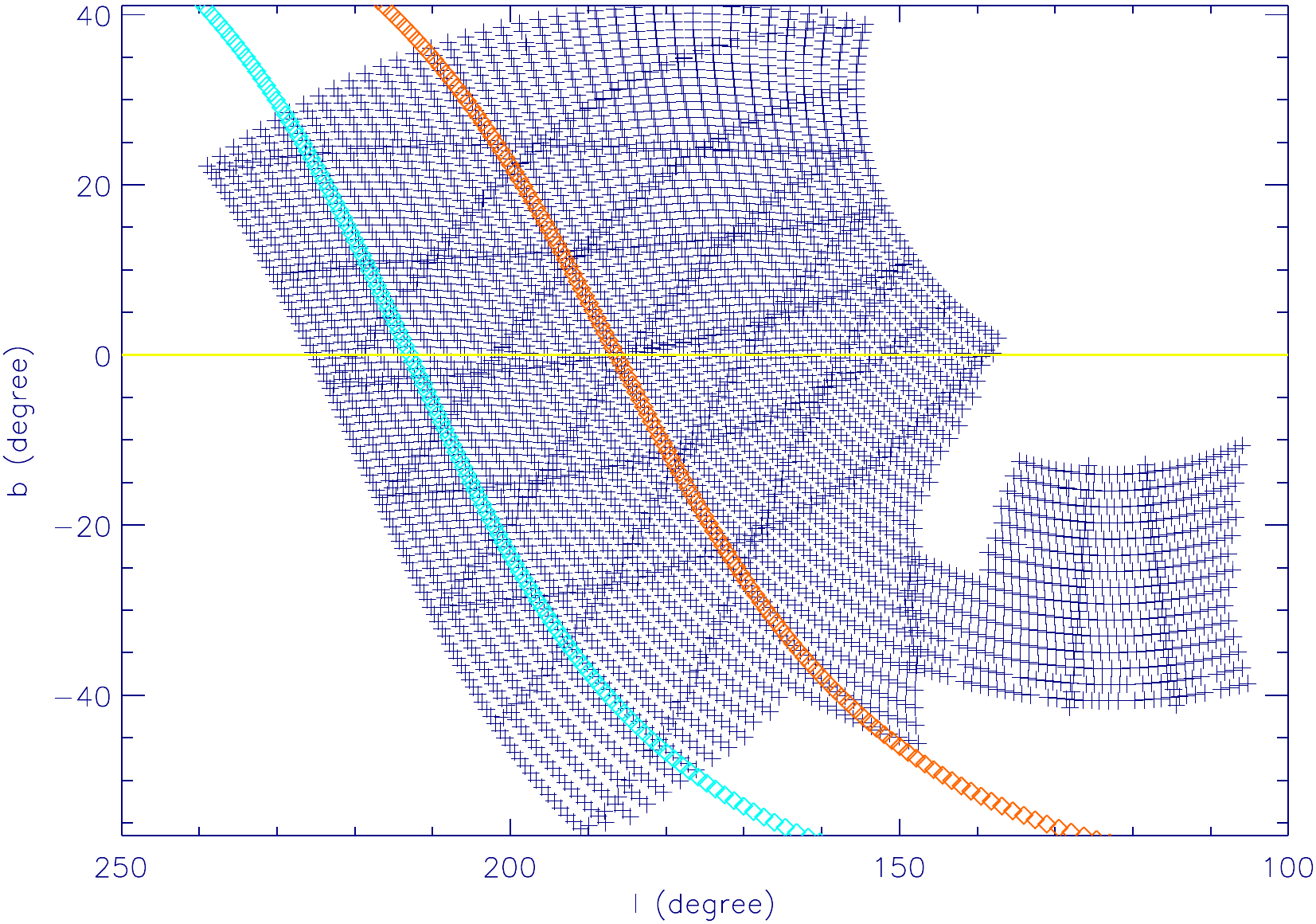}
\caption{{\it Left}: A small portion of the map of the artificial night light from NASA's "Blue Marble" pictures (http://www.blue-marble.de/; 2010 version), centered on the Xuyi site marked by an arrow. The large patch of heavy light pollution near the bottom is NanJing city and the one near the top is Huai'an city are marked. The approximate range scale is labeled in the left-corner. {\it Right}: The sky coverage of XSTPS-GAC fields. The yellow line denotes the Galactic plane, whereas the orange diamonds denote the ecliptic and celestial equator, respectively.}
\end{figure}

The sources for the night sky brightness include the diffuse stellar light
scattered by the interstellar medium, the solar and lunar light scattered by
atmospheric dust grains, the faint unresolved stars and galaxies within the
field of view, and the scattered artificial light of cities nearby. A nice
summary of the contributions of individual sources to the night sky brightness
of the La Palma site is published in 
\cite{Been2007}. 
We refer the reader to
this work for details of the origins and contributions of various light sources
and their potential effects on observations of different types of target. In
this work we measure the night sky brightness of the Xuyi site. 

\subsection{Methods}

The Xuyi site is just about a hundred kilometers away from municipal
NanJing city and Huai'An city, and is about tens of kilometers away from the nearest
Xuyi town. This could be seen on the map of the East Asia artificial night
light picture from NASA's "Blue Marble" project (2010 version), with a small
part near observation site was cut out and shown in the left panel of Fig.\,2. 
The artificial light pollution from the nearby cities and towns are
clearly visible and poses a danger to the site. In photometric nights, the
light pollution is less of a problem, but in partially cloudy nights even the
slightest light pollution may cause a big problem to astronomical observations
due to the effects of scattering. 

Only images of good quality collected in the XSTPS-GAC survey have been used to
measure the \nsb. Images collected under cloudy conditions can be easily
rejected from their abnormally bright background. To the limiting magnitude of
the survey, even for crowded fields near the Galactic plane, only a
small fraction of pixels are on stars and galaxies, while the rest fall on the
blank sky. Thus after clipping bright pixels of stars and galaxies, the
median count of blank pixels within the central 1024$\times$1024 CCD chip,
denoted $sky_{count}$, serves as a robust measurement of the sky background. Then,
the sky instrument brightness $sky_{inst}$ of a given image is calculated using: 
\begin{equation}
sky_{inst}=25.0-2.5 \times log_{10} (sky_{count}/scale^2)
\end{equation}  
where $scale$ is the angular size of a CCD pixel projected on the sky, i.e.
1$^{\rm ''}$.705 in this work.

\begin{figure}
\centering
\includegraphics[width=130mm]{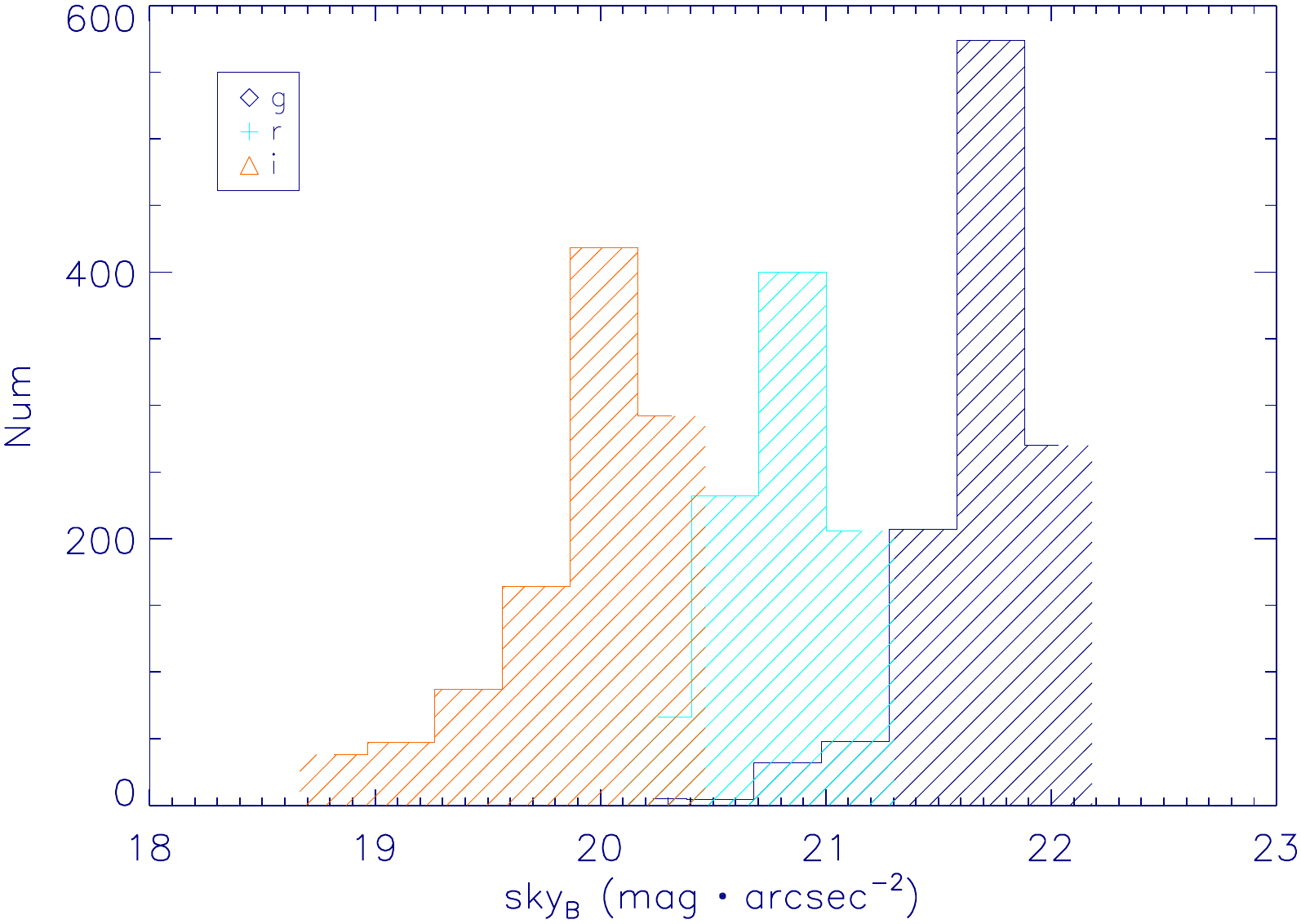}
\caption{Histograms of the night sky brightness in $g$~(blue), $r$~(cyan) and $i$~(orange) bands, deduced from images of Galactic latitudes $|b| >$ 15$^{\rm o}$, and taken under lunar phases $<$ 7 and angular distances $>$ 90$^{\rm o}$, more details could be seen in the text.}
\end{figure}

Then the sky brightness $sky_{B}$, in the units of mag/arcsec$^2$, can be calculated by:
\begin{equation}
sky_{B} = sky_{inst} + zp
\end{equation}
\begin{equation}
zp = \delta + ext + c_c
\end{equation}
where $zp$ is the instrument zero point, $ext$ is the atmospheric extinction which adopted 
same values as deduced in previous Section, $C_c$
is the color correction factor and $\delta$ is the zero point for the image.  Values
of $\delta$ in this work are from Yuan \etal. (in preparation).
Here the intrinsic \nsb~out of the atmosphere is calculated, and we find 
 a significant fraction of the
\nsb~at Xuyi site is from light pollution, as shown later. 

\subsection{Discussions}

The right panel of Fig.\,2 shows that XSTPS-GAC covers the Galactic disk and
the ecliptic equator, so the data can be used to study the effects of Galactic
disk and Zodiac light on the night sky brightness. Similarly, we can study the
dependence of the sky brightness on the solar cycle as well as the lunar phase. 

\begin{figure}
\centering
\includegraphics[width=120mm]{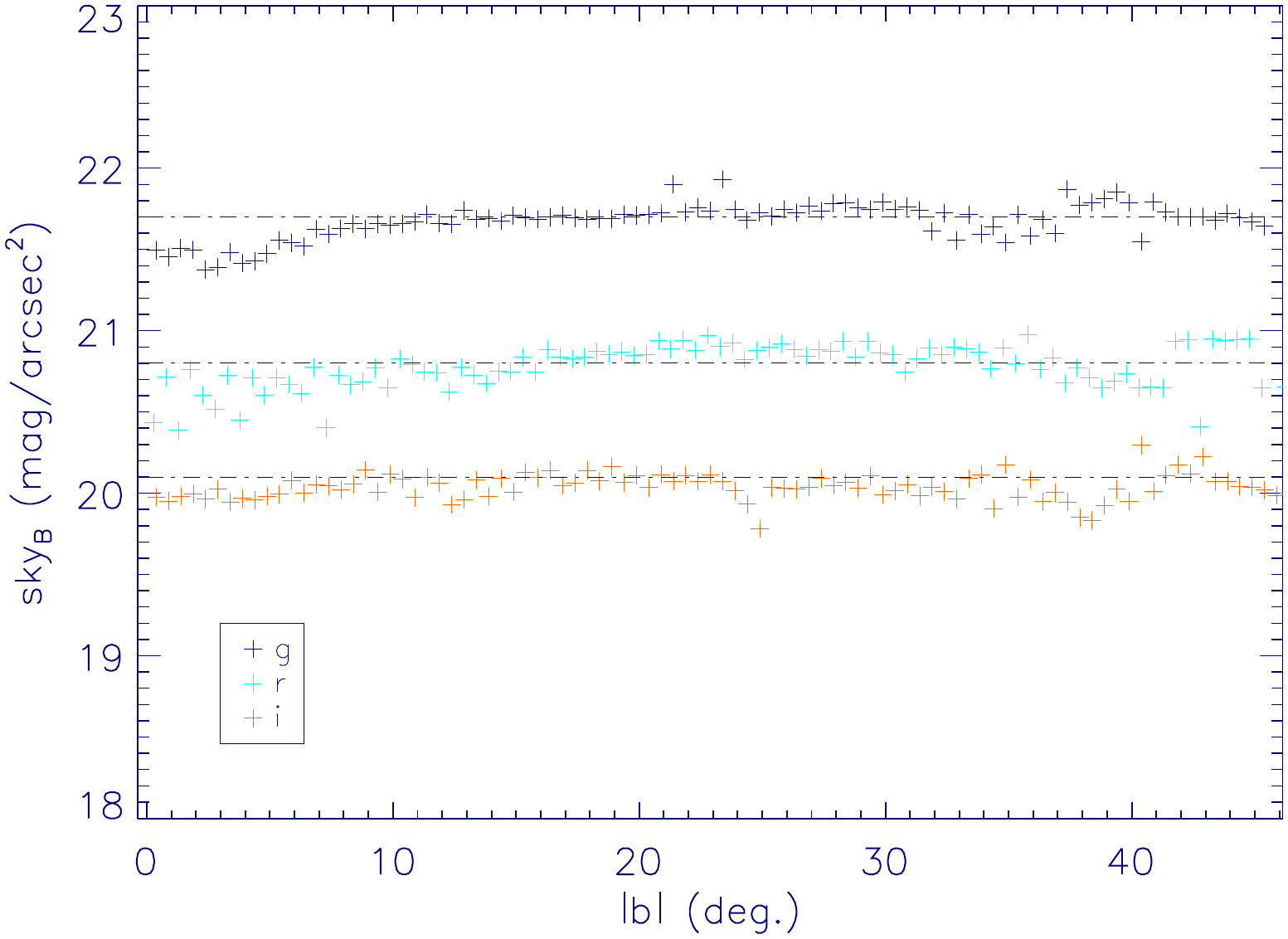}
\caption{The night sky brightness variations as a function of Galactic latitude. The dot-dash lines represent the median values adopted for the Xuyi site. Each plus point represents the median sky brightness with a binsize=0.5\,deg., in the units of mag/arcsec$^2$.}
\end{figure}

The XSTPS-GAC was carried out during the early phase of the 24th solar cycle,
from October 2009 to March 2012. The level of solar activity was normal and the
data showed no obvious evidence that the night sky brightness was affected by
the solar activities. This also implies that some caution must be exercised
when comparing the night sky brightness measurements presented in the current
work to other years of high levels of solar activities.

To minimize the effects of the bright Galactic disk, we calculated the night
sky brightness for each band using images of ecliptic latitudes $|\beta| >
$ 20$^{\rm o}$, Galactic latitudes $|b| >$ 15$^{\rm o}$, taken under lunar
phases $<$ 7, lunar angular distances $>$ 90$^{\rm o}$ and lunar altitude $<$ 0
degree. Fig.\,3 shows the histograms of \nsb~distribution that were deduced for the
$g$, $r$ and $i$~band. The median values of the \nsb~are $g$ = 21.7, $r$ =
20.8 and $i$ = 20.0
mag/arcsec$^2$. Fig.\,3 also shows that under
best observing conditions, the \nsb~can be as faint as 22.1, 21.2
and 20.4 mag/arcsec$^2$ in the $g$, $r$ and $i$ band, respectively. Based on the
transformation equations for population I stars between SDSS magnitudes and
$UBVR_CI_C$ given by 
\cite{Jordi2006} 
\begin{equation}
V-g = (-0.573 \pm 0.002)\times(g-r) - (0.016 \pm 0.002)
\end{equation}
and adopting $g-r = 0.8$ as the color of the night sky, we find that 
a typical value of \nsb~in $V$ band is $V = 21.2$ mag/arcsec$^2$, which is very
close to that of the Xinglong Station 
\citep{yaosRAA2012}, but can
reach to $V = 21.6$ mag/arcsec$^2$ under best conditions. Note that the \nsb~at
the world-class sites can reach a typical value of $V = 21.9$ mag/arcsec$^2$ 
\citep{Benn1998},
suggesting that a significant fraction of the \nsb~at the Xuyi site 
is from light pollution.

To investigate the effect of diffuse light from the Galactic disk,
\nsb~deduced from images of ecliptic latitudes $|\beta| >$ 15$^{\rm o}$, secured under
lunar phases $P <$  7, lunar altitudes $<$ 0 and lunar angle distances $>$
90$^{\rm o}$ are plotted against Galactic lattitudes in Fig.\,4. Even though there
are some fluctuations, we can see that the Galactic disk enhances the night
sky by about 0.3 and 0.2 mag/arcsec$^2$ in the $g$~and $r$~bands, respectively.
However, the effect in the $i$~band is not obvious. 
Similarly, images of Galactic latitudes $|b| >$ 10$^{\rm o}$, obtained
under lunar phases $P <$ 7, lunar altitudes $<$ 0, and lunar angle distances
$>$ 90$^{\rm o}$ were used to study the variations of the \nsb~with ecliptic
latitude. No obvious variations of the night sky brightness with ecliptic latitudes are found
in all the three bands.

\begin{figure}
\centering
\includegraphics[width=144mm]{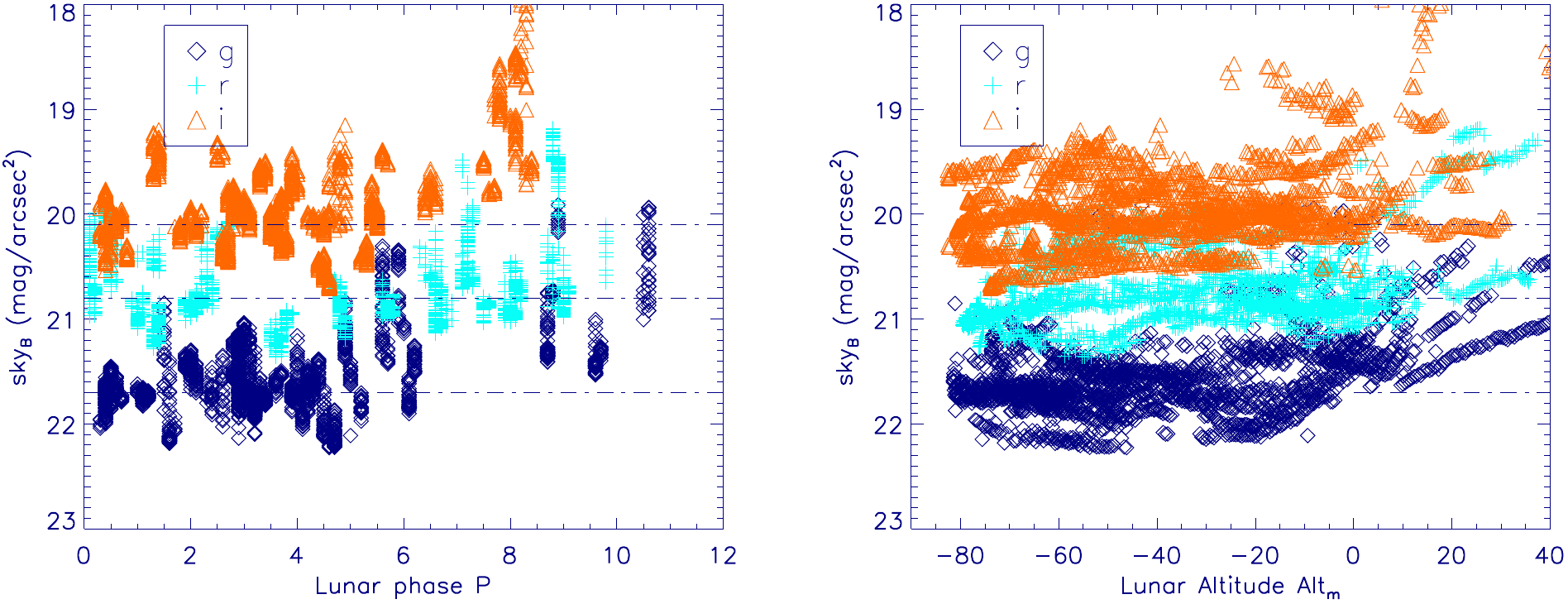}
\includegraphics[width=70mm]{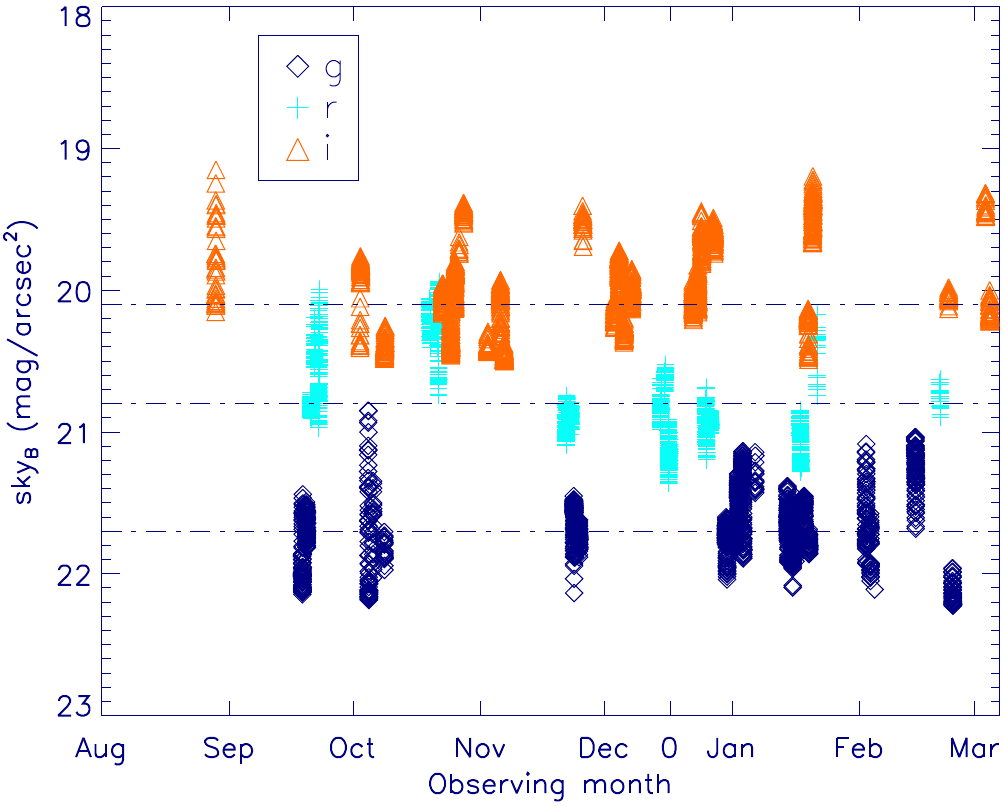}
\caption{Variations of the night sky brightness with lunar phase $P$ (top-left), altitude alt$_{\rm m}$ (top-right). 
In the bottom panel, only data points of $P <$ 6 and alt$_{\rm m} <$ 0 are shown. The color codes of the data points are the same as in Fig.\,3.}
\end{figure}

Finally, we study the dependence of night sky brightness on the lunar phase $P$,
lunar altitude alt$_{\rm m}$, and the angular distance $D$ between the field
center and the Moon. A similar work on the effects of the moon on the night sky
brightness is presented by \cite{Kris1991}. 
The lunar phase $P$ is normalized to $0$ -- $15$, where $0$ represents new moon
and $15$ full moon. Note all the observations were taken under lunar phase $P
<$ 10. Our results are shown in Fig.\,5. The top left panel of Fig.\,5 shows that the night sky
brightness is nearly flat for $P <$ 6, but increases significantly
thereafter at a rate of about 0.3, 0.2 and 0.2 mag/arcsec$^2$ per night for the
$g$, $r$ and $i$~band, respectively. The night sky brightness is also found to
be nearly constant for lunar altitudes smaller than 0 degree, i.e. below the
horizon, and then brightens by about 0.2, 0.1 and 0.1 mag/arcsec$^2$ every 10
deg.~for the three bands respectively as the moon rises. 
The bottom panel of Fig.\,5 shows the seasonal variations of the night sky
brightness, only images of $P <$ 6 and alt$_{\rm m} <$ 0 were used. 
No clear seasonal variations are seen in the three bands.  

\section{Conclusions}
\label{sect:con}

In this work, we present measurements of the optical broadband \aec~and the
\nsb~at the PMO Xuyi site, based on the large data set collected for the
DSS-GAC survey from 2009 to 2011. The mean extinction coefficients are 0.69,
0.55 and 0.38 mag/airmass for the $g$, $r$~and $i$~band, respectively. The values  
are larger than the best astronomical sites, and vary from night 
to night. So, this result here just represents the typical atmospheric extinction 
for the site, it's unwise to use a single mean extinction coefficient to do the 
flux calibration for the whole data, which will introduce large errors.

The typical night sky brightness are 21.7, 20.8 and 20.0 mag/arcsec$^{\rm 2}$
in the $g$, $r$~and $i$~band, respectively, which corresponds to $V = 21.2$. 
The \nsb~could reach as faint as 22.1, 21.2 and 20.4 mag/arcsec$^{\rm 2}$ under best conditions.
A significant fraction of night sky is from light pollution. The diffuse light from 
the Galactic disk enhances the \nsb~ by about 0.3, 0.2 and 0.2 mag/arcsec$^{\rm 2}$ for the 
 $g$, $r$~and $i$~band, respectively. But the Zodiac light has almost no obvious 
influences. There is no variations if lunar phase $P <$~6 and lunar altitudes alt$_m < $0.  
The \nsb~will be respectively brightened by about 0.3, 0.2 and 0.2\,mag per night 
after lunar phase $P >$~7 and 0.02, 0.01 and 0.01\,mag per degree with lunar altitudes 
alt$_m$~above the horizon. 
The relatively faint night sky brightness makes it a good observational site in 
China for wide field surveys. To protect the site from further deterioration, 
light pollutions must be strictly controlled for any resort development around the area.

\normalem
\begin{acknowledgements}
This work is supported by the National Natural Science Foundation of China (NSFC) grant \#11078006 and \#10933004. And this work is also supported by the Minor Planet Foundation of Purple Mountain Observatory.
\end{acknowledgements}

\bibliographystyle{raa}
\bibliography{extin.bib}

\label{lastpage}

\end{document}